\documentclass{elsart}
\usepackage[dvips]{graphicx}
\def\simlt{\mathrel{\hbox to 0pt{\lower 3.5pt\hbox{$\mathchar"218$}\hss}
      \raise 1.5pt\hbox{$\mathchar"13C$}}}
\def\simgt{\mathrel{\hbox to 0pt{\lower 3.5pt\hbox{$\mathchar"218$}\hss}
      \raise 1.5pt\hbox{$\mathchar"13E$}}}
\def\cl{C_\ell}
\def\etal{{\it et al. }}
\def\deg{^{\circ}}
\def\bepsilon{\mbox{\boldmath $\epsilon$}}
\begin{document}
\runauthor{Barreiro}
\begin{frontmatter}
\title{The Cosmic Microwave Background: State of the Art}
\author[camb,ifca,dpto]{R. Bel\'en Barreiro}
\address[camb]{Astrophysics Group, Cavendish Laboratory, Madingley Road
\\Cambridge CB3 0HE, UK}
\address[ifca]{Instituto de F\'\i sica de Cantabria, 
Facultad de Ciencias, Avda. de los Castros s/n \\ 39005 Santander, Spain}
\address[dpto]{Depto. F\'\i sica Moderna, Facultad de Ciencias,
Avda. de los Castros s/n \\ 39005 Santander, Spain}
\begin{abstract}
We review the current status of the cosmic microwave
background (CMB) radiation, including a brief discussion of some basic 
theoretical aspects as well as a summary of anisotropy detections and CMB 
experiments. We focus on the description of some relevant characteristics 
of the microwave foregrounds, on the discussion of the different estimators 
proposed in the literature to detect non-Gaussianity and on outlining
the bases of different reconstruction methods that have been 
applied to the CMB.
\end{abstract}
\end{frontmatter}

\section{Introduction}
The Cosmic Microwave Background (CMB) Radiation constitutes  
one of the most powerful tools of Cosmology.
This radiation is a relic from a hot and dense past of the universe, 
produced at the Big Bang and freely propagated
$\sim 300000$ years after it. 
Before these early times, due to the high temperature, 
matter is completely ionized. Compton scattering tightly couples 
the photons to the electrons which are in turn coupled to the baryons
by electromagnetic interactions. 
As the universe expands, the temperature decreases and at a 
redshift $z \sim 1000$, T has dropped to $\sim 3000 K$,
allowing free electrons and protons to form neutral atoms.
At this time, known as {\it decoupling}, the universe becomes 
transparent, the photons are last scattered
off by the electrons and can freely propagate, giving rise to
the CMB.
Due to the thermal equilibrium between matter and radiation before 
decoupling and their lack of interaction after that time, the CMB exhibits 
a blackbody spectrum with a present temperature of $T_o = 2.73 K$.
The temperature of the CMB has dropped since decoupling due to
the expansion of the universe according to 
$T(z) = (1+z)T_o,$ where $T(z)$ denotes the temperature measured 
by an observer at redshift $z$.

The existence of the CMB was first predicted by Gamow and his 
collaborators in 1948, 
when studying the light-element synthesis in the primordial universe.
They predicted that this relic radiation should still be ubiquitous
today, with a temperature of about 5K (Gamow 1948a,b, Alpher \&
Herman 1948).
It was not until 1964, when the discovery of the CMB was made by Penzias 
and Wilson (published in 1965). 
They detected an excess of noise in their horn 
antenna with a temperature $\sim 3K$ coming 
from all directions in the sky and being very uniform.
For a historical introduction see Partridge (1995).
The temperature of the CMB has been measured by the FIRAS 
instrument on board the COBE satellite
to be $T_o=2.728 \pm 0.004$K (Fixsen \etal 1996).
The prediction and the subsequent detection of the CMB is one of the 
strongest supports for the Big Bang model.

Since the CMB freely propagated after
the decoupling time, it carries
information about how the universe was at $z \simlt 1000 $.
The fact that the CMB is very homogeneous, means that so was the primitive 
universe. 
However, the matter in our universe clusters on a wide range of scales,
forming all the structures we see today.
If all these structures were formed
via gravitational instability, those density fluctuations should
already be present at early times, leaving their imprint as
temperature anisotropies in the CMB.

\section{Temperature anisotropies}

The temperature anisotropies of the CMB are described by a 2-dimensional random
field $\frac{\Delta T}{T}(\vec{n}) \equiv \frac{T(\vec{n})-T_o}{T_o}$,
where $\vec{n}$ is a unit vector on the sphere.
It is usual to expand the field in spherical harmonics:
\begin{equation}	
\frac{\Delta T}{T}(\vec{n}) = 
\sum_{\ell =1}^{\infty}\sum_{m=-\ell}^{\ell}a_{\ell m}Y_{\ell m}(\vec{n}) \, ,
\end{equation}
In this expansion, low $\ell$'s correspond to anisotropies on large
angular scales whereas large $\ell$'s reflect the anisotropies 
at small scales.
The $a_{\ell m}$ coefficients are independent random variables of 
mean $< a_{\ell m} > = 0$. If the temperature fluctuations are
{\it statistically} isotropic, the variance of the $a_{\ell m}$ 
coefficients is independent of $m$:
\begin{equation}
<a_{\ell m} a_{\ell' m'}^*> = C_{\ell}\delta_{\ell \ell'}\delta_{mm'} 
\, ,
\end{equation}
where the averages are to be taken over statistical ensembles.
The set of $C_\ell$'s constitutes the {\it angular power spectrum}.
In the case of Gaussian fluctuations, as predicted by inflation,
the 2-point correlation function $C(\theta)$ completely 
characterizes the random temperature field and can be written as:
\begin{equation}
C(\theta)=\left<\left( \frac{\Delta T}{T}\left(\vec{n_1}\right)
\cdot\frac{\Delta T}{T}\left(\vec{n_2}\right)\right)\right> = 
\sum_{\ell}\frac{(2\ell+1)}{4\pi}\cl P_\ell\left(cos\theta\right)
\, ,
\end{equation}
where $P_\ell$ is the Legendre polynomial of order $\ell$ and
$\theta$ is the angle formed by the vectors $\vec{n_1}$ and $\vec{n_2}$
on the sky. Therefore, for standard inflationary models,
the angular power spectrum contains all the statistical information
about the field, being the fundamental quantity in the
theory of the CMB anisotropies.
The $\cl$'s can be accurately calculated for the inflationary
models as a function of the cosmological parameters 
(e.g. Seljak \& Zaldarriaga 1996, Hu \etal 1998). 
Thus, accurate measurements of the angular power spectrum would provide 
tight constrains on the cosmological model (Bond \etal 1997). 
For theories generating non-Gaussian fluctuations, study of higher order 
moments becomes necessary. However, the angular power spectrum is still a 
fundamental test of the viability of those theories.

The accuracy with which a given $\cl$ can be measured is limited by
the so-called {\it cosmic variance}, which is due to the fact
of having just a single realization of the temperature field, our universe. 
For Gaussian fluctuations each $\cl$ is drawn from a $\chi^2$ distribution
with $(2\ell+1)$ degrees of freedom. The minimum variance of a 
measured $\cl$ is given by $[2/(2\ell+1)]C_\ell^2$, mainly affecting 
the low $\ell$'s (large scales).
Another effect that reduces our ability to accurately measure
the angular power spectrum is the {\it sample variance},
which is due to partial coverage of the sky.
This effect enhances the cosmic variance by a factor $\simeq 4\pi/A$ 
where $A$ is the solid angle covered by the experiment
(Scott \etal 1994). Even full-sky coverage satellite missions can be
affected by sample variance, since we may need to discard 
highly contaminated parts of the data 
(such as the Galactic plane) when studying the CMB.
It must be pointed out that cosmic and sample variance are present 
independently of the resolution and sensitivity of the experiment.

\section{Summary of anisotropy detections}

\begin{figure}[!t]
\resizebox{\hsize}{!}{\includegraphics[angle=270]
{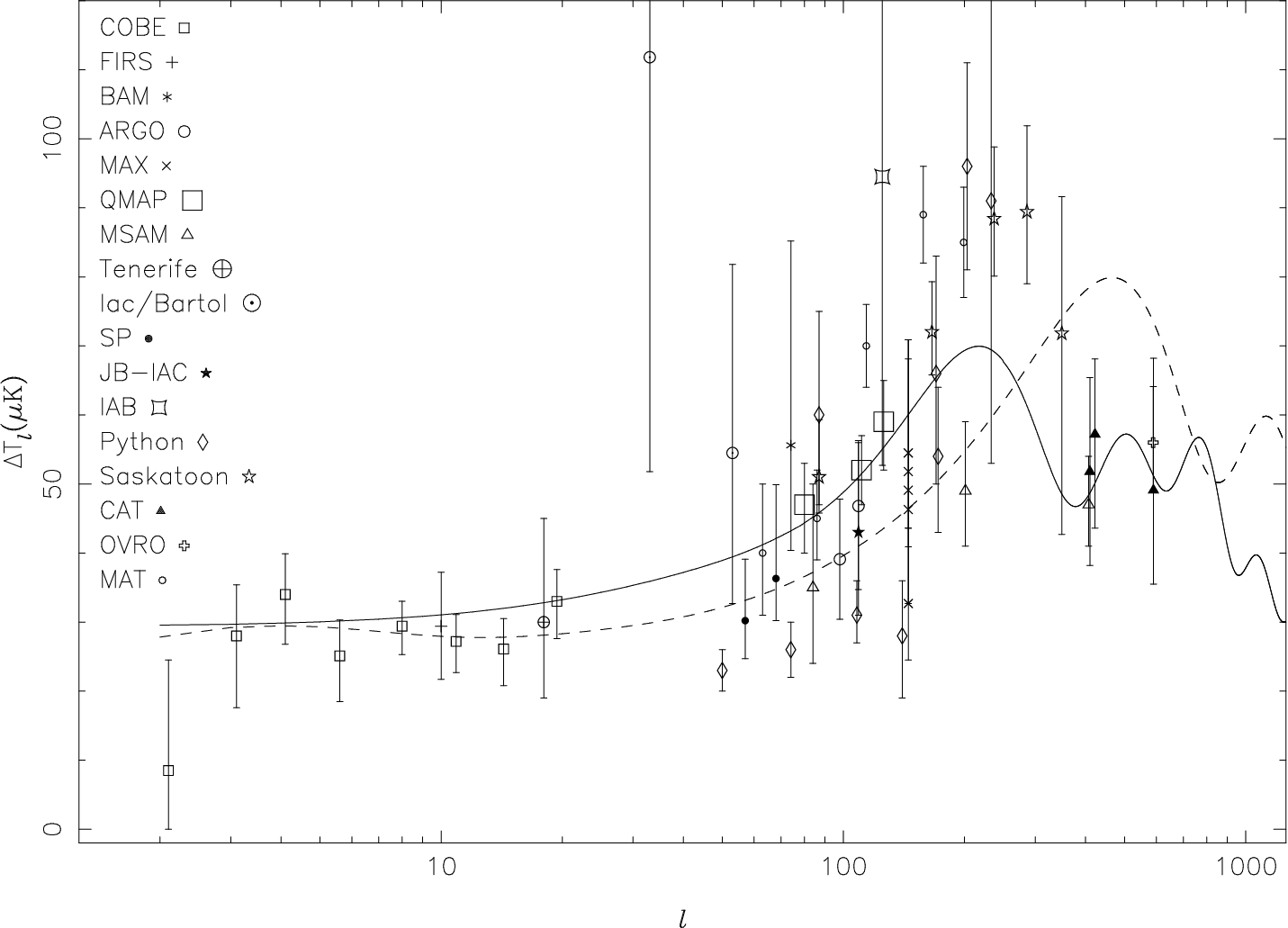}}
\caption[Anisotropy detections in the CMB]
{Anisotropy detections in the CMB with the error bars showing
the $1\sigma$ confidence level. The solid and dashed line 
correspond to the angular power spectrum predicted for a standard 
flat and open ($\Omega=0.3)$ CDM models, respectively.}
\label{datos_exp}
\end{figure}

\begin{table}[t]
\caption[Anisotropy detections I]{Summary of anisotropy detections by COBE and 
balloon-borne experiments}
\label{det_bal}
\begin{center}
\begin{tabular}{|l|c|c|c|c|l|}
\hline
Experiment & ${\Delta T_\ell}^{+\sigma}_{-\sigma} (\mu K) $ & $\ell_{min}$ & 
$\ell_{eff}$ & $\ell_{max}$ & Reference\\
\hline
COBE 1 & $8.5^{+16.0}_{-8.5}$ & 2 & 2.1 & 2.5 & Tegmark \& Hamilton 1997 \\
COBE 2 & $28.0^{+7.4}_{-10.4}$ & 2.5 & 3.1 & 3.7 & Tegmark \& Hamilton 1997 \\
COBE 3 & $34.0^{+5.9}_{-7.2}$ & 3.4 & 4.1 & 4.8 & Tegmark \& Hamilton 1997 \\
COBE 4 & $25.1^{+5.2}_{-6.6}$ & 4.7 & 5.6 & 6.6 & Tegmark \& Hamilton 1997 \\
COBE 5 & $29.4^{+3.6}_{-4.1}$ & 6.8 & 8.0 & 9.3 & Tegmark \& Hamilton 1997 \\
COBE 6 & $27.7^{+3.9}_{-4.5}$ & 9.7 & 10.9 & 12.2 & Tegmark \& Hamilton 1997 \\
COBE 7 & $26.1^{+4.4}_{-5.3}$ & 12.8 & 14.3 & 15.7 & Tegmark \& Hamilton 1997 \\
COBE 8 & $33.0^{+4.6}_{-5.4}$ & 16.6 & 19.4 & 22.1 & Tegmark \& Hamilton 1997 \\
\hline
FIRS & $29.4^{+7.8}_{-7.7}$ & 3 & 10 & 30 & Ganga \etal 1994 \\
BAM & $55.6^{+29.6}_{-15.2}$ & 28 & 74 & 97 & Tucker \etal 1997 \\
ARGO 1& $39.1^{+8.7}_{-8.7}$ & 52& 98 & 176 & 
de Bernardis \etal 1994\\
ARGO 2 & $46.8^{+9.5}_{-12.1}$ & 53 & 109 & 179 & Masi \etal 1996\\
MAX GUM & $54.5^{+16.4}_{-10.9}$ & 78 & 145 & 263 & Tanaka \etal 1996 \\
MAX ID & $46.3^{+21.8}_{-13.6}$ & 78 & 145 & 263 & Tanaka \etal 1996 \\
MAX SH & $49.1^{+21.8}_{-16.4}$ & 78 & 145 & 263 & Tanaka \etal 1996 \\
MAX HR & $32.7^{+10.9}_{-8.2}$ & 78 & 145 & 263 & Tanaka \etal 1996 \\
MAX PH & $51.8^{+19.1}_{-10.9}$ & 78 & 145 & 263 & Tanaka \etal 1996 \\
QMAP I+II  & $47^{+6}_{-7}$ & 39 & 80 & 121 & 
de Oliveira-Costa \etal 1998 \\
QMAP I+II  & $59^{+6}_{-7}$ & 72 & 126 & 180 & de Oliveira-Costa \etal 1998 \\
QMAP I+II  & $52^{+5}_{-5}$ & 47 & 111 & 175 & de Oliveira-Costa \etal 1998 \\
MSAM I  & $35^{+15}_{-11}$ & 39 & 84 & 130 & Wilson \etal 1999 \\
MSAM I  & $49^{+10}_{-8}$ & 131 & 201 & 283 & Wilson \etal 1999 \\
MSAM I  & $47^{+7}_{-6}$ & 284 & 407 & 453 & Wilson \etal 1999 \\
\hline
\end{tabular}
\end{center}
\end{table}
\begin{table}[!p]
\caption[Anisotropy detections II]
{Summary of anisotropy detections by ground-based experiments}
\label{det_ground}
\begin{center}
\begin{tabular}{|l|c|c|c|c|l|}
\hline
Experiment & ${\Delta T_\ell}^{+\sigma}_{-\sigma} (\mu K) $ & $\ell_{min}$ & 
$\ell_{eff}$ & $\ell_{max}$ & Reference\\
\hline
Tenerife & $30^{+15}_{-11}$ & 11 & 18 & 27 & Guti\'errez \etal 1999\\
IAC/Bartol 1 & $111.8^{+65.4}_{-60.0}$ & 20 & 33 & 57 & Femenia \etal 1998 \\
IAC/Bartol 2 & $54.5^{+27.3}_{-21.8}$ & 38 & 53 & 75 & Femenia \etal 1998 \\
SP91 & $30.2^{+8.9}_{-5.5}$& 31 & 57 & 106 & Gundersen \etal 1995 \\
SP94 & $36.3^{+13.6}_{-6.1}$& 36 & 68 & 106 & Gundersen \etal 1995 \\
JB-IAC & $43^{+13}_{-12}$ & 90 & 109 & 128 & Dicker \etal 1999 \\
IAB & $94.5^{+41.8}_{-41.8}$ & 60 & 125 & 205 & Piccirillo \& Calisse 1993 \\
Python III & $60^{+15}_{-13}$ & 49 & 87& 105 & Platt \etal 1997\\
Python I+II+III & $66^{+17}_{-16}$ & 120 & 170 & 239 & Platt \etal 1997\\
Python V & $23^{+3}_{-3}$ & 21 & 50 & 94 & Coble \etal 1999 \\
Python V & $26^{+4}_{-4}$ & 35 & 74 & 130 & Coble \etal 1999 \\
Python V & $31^{+5}_{-4}$ & 67 & 108 & 157 & Coble \etal 1999 \\
Python V & $28^{+8}_{-9}$ & 99 & 140 & 195 & Coble \etal 1999 \\
Python V & $54^{+10}_{-11}$ & 132 & 172 & 215 & Coble \etal 1999 \\
Python V & $96^{+15}_{-15}$ & 164 & 203 & 244 & Coble \etal 1999 \\
Python V & $91^{+32}_{-38}$ & 195 & 233 & 273 & Coble \etal 1999 \\
MAT & $40^{+10}_{-9}$ & 45 & 63 & 81 & Torbet \etal 1999 \\ 
MAT & $45^{+7}_{-6}$ & 64 & 86 & 102 & Torbet \etal 1999 \\
MAT & $70^{+6}_{-6}$ & 90 & 114 & 134 & Torbet \etal 1999 \\
MAT & $89^{+7}_{-7}$ & 135 & 158 & 180 & Torbet \etal 1999 \\
MAT & $85^{+8}_{-8}$ & 170 & 199 & 237 & Torbet \etal 1999 \\
Saskatoon 1 ($^*$) & $51.0^{+8.3}_{-5.2}$ & 58 & 87 & 126 & 
Netterfield \etal 1997 \\
Saskatoon 2 & $72.0^{+7.3}_{-6.2}$ & 123 & 166 & 196 & Netterfield \etal 1997\\
Saskatoon 3 & $88.4^{+10.4}_{-8.3}$ & 196 & 237 & 266 
& Netterfield \etal 1997\\
Saskatoon 4 & $89.4^{+12.5}_{-10.4}$ & 248 & 286 & 310 
& Netterfield \etal 1997\\
Saskatoon 5 & $71.8^{+19.8}_{-29.1}$ & 308 & 349 & 393 & 
Netterfield \etal 1997\\
CAT 1 & $51.8^{+13.6}_{-13.6}$ & 339 & 410 & 483 & Scott \etal 1996 \\
CAT 1 & $49.1^{+19.1}_{-13.6}$ & 546 & 590 & 722 & Scott \etal 1996 \\
CAT 2 & $57.2^{+10.9}_{-13.6}$ & & 422 & & Baker \etal 1999\\
OVRO & $56^{+8.1}_{-6.9}$ & 361 & 589 & 756 & Leitch \etal 1998 \\ 
\hline
\end{tabular}
\begin{itemize}
\item [$\ast$] The Saskatoon data include the latest calibration
correction (Leitch \etal 1998).
\end{itemize}
\end{center}
\end{table}

Before COBE, the only temperature anisotropy detected in the CMB
was of dipolar nature (Smoot \etal 1977).
This dipolar component is the largest anisotropy present in the CMB and it 
is due to a Doppler shift caused by 
the motion of the observer with respect to the rest frame of the CMB
(i.e., it has an extrinsic origin):
\begin{equation}
T(\theta) \approx T_o(1+(v/c)\cos\theta) \,\, , v/c << 1 \, , 
\end{equation}
where $v$ is the velocity of the observer with respect to the CMB 
and $\theta$ the angle formed by the line of sight and the velocity.
The COBE team found an amplitude for the dipole of $3.372 \pm 0.007$mK
with a maximum in the direction $(\ell,b)=(264.14^\circ \pm
0.30^\circ, 48.26\deg \pm 0.30\deg)$ (Fixsen \etal 1996).

In 1992, the COBE team announced the first detection of
intrinsic anisotropy of the CMB, at angular scales of $\sim 10^{\circ}$,
at the level of $\sim 10^{-5}$ (Smoot \etal 1992).
Since then, more than a dozen of groups have reported 
anisotropy detections 
spanning over many angular scales.

When observing the microwave sky, we must take into account the
resolution, sensitivity and observing technique of the experiment.
The sensitivity of the experiment to any given scale is defined
by the window function $W_\ell$ (e.g. White \& Srednicki 1994, 
Cay\'on 1996).
For instance, for a Gaussian beam the window function is 
$W_\ell=e^{-\ell(\ell+1)^2\sigma_b^2}$, where $\sigma_b$ is the 
dispersion of the beam.
In addition, instrumental noise must be taken into account when
interpreting the data.
The temperature fluctuation averaged over the sky observed by an
experiment is given by:
\begin{equation}
\left(\frac{\Delta T}{T}\right)_{rms}^2=\sum_\ell\frac{2\ell+1}{4\pi}
\cl W_\ell 
\end{equation}

In order to compare the measured fluctuations from different experiments
with the angular power spectrum predicted by the theory, the 
power per logarithmic scale 
$\left(\frac{\Delta T}{T}\right)^2_\ell \equiv 
\ell(\ell+1)\cl/(2\pi)$ is commonly used.
Assuming that this quantity is flat over the range of multipoles
where the experiment is more sensitive, it can be easily 
estimated from the measured temperature
fluctuation $\left(\frac{\Delta T}{T}\right)_{rms}$ as
(Bond 1995):
\begin{equation}
\left(\frac{\Delta T}{T}\right)^2_\ell =
\frac{\left( \frac{\Delta T}{T} \right)_{rms}^2}{\sum_\ell W_\ell 
\frac{2\ell+1}{2\ell(\ell+1)}}
\, ,
\end{equation}
what is usually known as the {\it band power}.

In figure~\ref{datos_exp}, a compilation of anisotropy 
detections is plotted,
where the vertical error bars show the $1\sigma$ confidence level.
Horizontal error bars, that account for the range of multipoles
where the experiment is sensitive, are not plotted for the sake of 
clarity.
The value of the detections and the $1\sigma$
error bars are listed in table~\ref{det_bal} (from COBE and balloon-borne
experiments) and~\ref{det_ground}  (from ground based experiments).
$\ell_{eff}$ corresponds to the centre of the window function and
$\ell_{min}$ and $\ell_{max}$ to the multipoles where it drops to half 
of its central value
(except for COBE, where $\ell_{min}$ and $\ell_{max}$ indicate the
rms width of the window function as calculated in Tegmark \& Hamilton
1997). Part of these numbers have been taken from the compilations of
Griffiths \etal (1999) and M.Tegmark's web 
page\footnote{http://www.sns.ias.edu/$\sim$max/\#CMB}.
For comparison, the angular power spectrum for two standard CDM models
($H_0=50$, $\Omega_b=0.05$ and initial scale invariant
perturbations) with $\Omega=1$ (solid line) and $\Omega=0.3$ 
are also plotted (dashed line).

At large scales ($\ell \simlt 20)$, the data are 
consistent with a Harrison-Zel'dovich primordial spectrum 
$\cl \propto 1/(\ell(\ell+1))$ as
predicted by inflation (see $\S$~\ref{initial perturbations}).
At medium angular scales, although the scatter is still large, 
the data seem to indicate the presence
of a Doppler peak at $\ell \sim 200$.
New data obtained from several experiments capable of measuring this
range of $\ell$'s with good precision, such as Boomerang or Maxima,
are currently under analysis and will confirm whether the first 
Doppler peak has been actually detected.

\section{Summary of CMB experiments}

\begin{table}[!t]
\caption{Summary of ground based experiments}
\label{exp_ground}
\begin{center}
\begin{tabular}{|l|c|c|c|}
\hline
Experiment$^{\dagger}$ & Resolution & Frequency (GHz) & Detectors \\
\hline
APACHE(c)  & $30'$ & 90-259  & Bol \\
ACBAR(p) & $\sim 5'$ & 150-450 & Bol \\ 
ATCA(c)$^{\ast}$   & $2'$ & 8.7  & HEMT \\
CAT(c)$^{\ast}$    & $10'-30'$ & 13-17  & HEMT \\
CBI(p)$^{\ast}$    & $4.5'-10'$  & 26-36  & HEMT \\
CG(c) & $\sim 1'$ & 1-32 & HEMT \\
DASI(p)$^{\ast}$ & $0.25\deg-1.15\deg$ & 26-36   & HEMT \\
HACME/SP(f)    & $46'$  & 39-43  & HEMT \\
IAC/Bartol(f) & $2\deg$  & 91-272   & Bol \\
JB-IAC(c)$^{\ast}$ & $2\deg$ & 33 & HEMT \\
MAT(c)  & $12'$  & 30-440  & HEMT/SIS \\
OVRO 40/5(c) & $7'-22'$ & 14.5-32  & HEMT \\ 
Python(f) & $45'$ & 37-90  & Bol/HEMT \\
Saskatoon(f) & $0.5\deg-1.5\deg$ & 26-46  & HEMT \\ 
SuZIE(c) & $\sim 2'$ & 143-350 & Bol \\
Tenerife(c) & $5\deg$ & 10-33  & HEMT \\
Viper(c) & $20'-2'$ & 35-400  & HEMT/Bol \\
VLA(f)$^{\ast}$ & $\sim 10''$ & 8.4  & HEMT \\
VSA(p)$^{\ast}$ & $0.25\deg-2\deg$ & 26-36   & HEMT \\
White Dish(f) & $12'$ & 90 & Bol \\
\hline
\end{tabular}
\end{center}
\begin{itemize}
\item [$\dag$] An `f' after the experiment's name means it's finished;
a `c' denotes current; a `p' denotes planned.
\item [$\ast$] Interferometer
\end{itemize}
\end{table}

\begin{table}[t]
\begin{center}
\caption{Summary of balloon-borne experiments}
\label{exp_balloon}
\begin{tabular}{|l|c|c|c|}
\hline
Experiment$^{\dagger}$ & Resolution & Frequency (GHz) & Detectors \\
\hline
ACE(p)          & $9'$ & 90 & HEMT \\
Archeops(p) & $8',5.5',5'$ & 143-353  & Bol \\
ARGO(f)  & $52'$ & 150-600 & Bol \\
BAM(c)     & $42'$ & 93-276  & Bol \\ 
BEAST(p)      & $25',19',9'$ & 30-90  & HEMT \\
BOOMERanG(c)   & $20',12'$ & 90-400  & Bol \\
FIRS(f)  & $3.8^{\circ}$ & 170-680  & Bol \\
MAX(f)    & $30'$  & 105-420  & Bol \\
MAXIMA(c)   & $10'$  & 150-410  & Bol \\
MSAM I(f)  & $37'$ & 170-680  & Bol \\
MSAM II(c) & $20'$ & 70-170  & Bol \\
QMAP(f) & $54',36'$ & 30-40   & HEMT \\ 
TopHat(c) & $20'$ & 70-630  & Bol \\ 
\hline
\end{tabular}
\end{center}
\begin{itemize}
\item [$\dagger$] An `f' after the experiment's name means it's finished;
a `c' denotes current; a `p' denotes planned.
\end{itemize}
\end{table}

During the last years, there has been an explosion of experiments 
dedicated to measure the CMB temperature anisotropies. 
Ground based, including interferometers, and balloon-borne experiments 
have been designed to probe a large range of angular scales.
Two satellite missions have also been approved: the 
MAP satellite of the NASA and the Planck Mission
of the ESA. The expected launch dates are 2001 and 2007, respectively. 
Both satellites will provide 
multifrequency full sky maps at unprecedented angular resolution and
sensitivity. 

\begin{figure}[!t]
\resizebox{12cm}{!}{\includegraphics{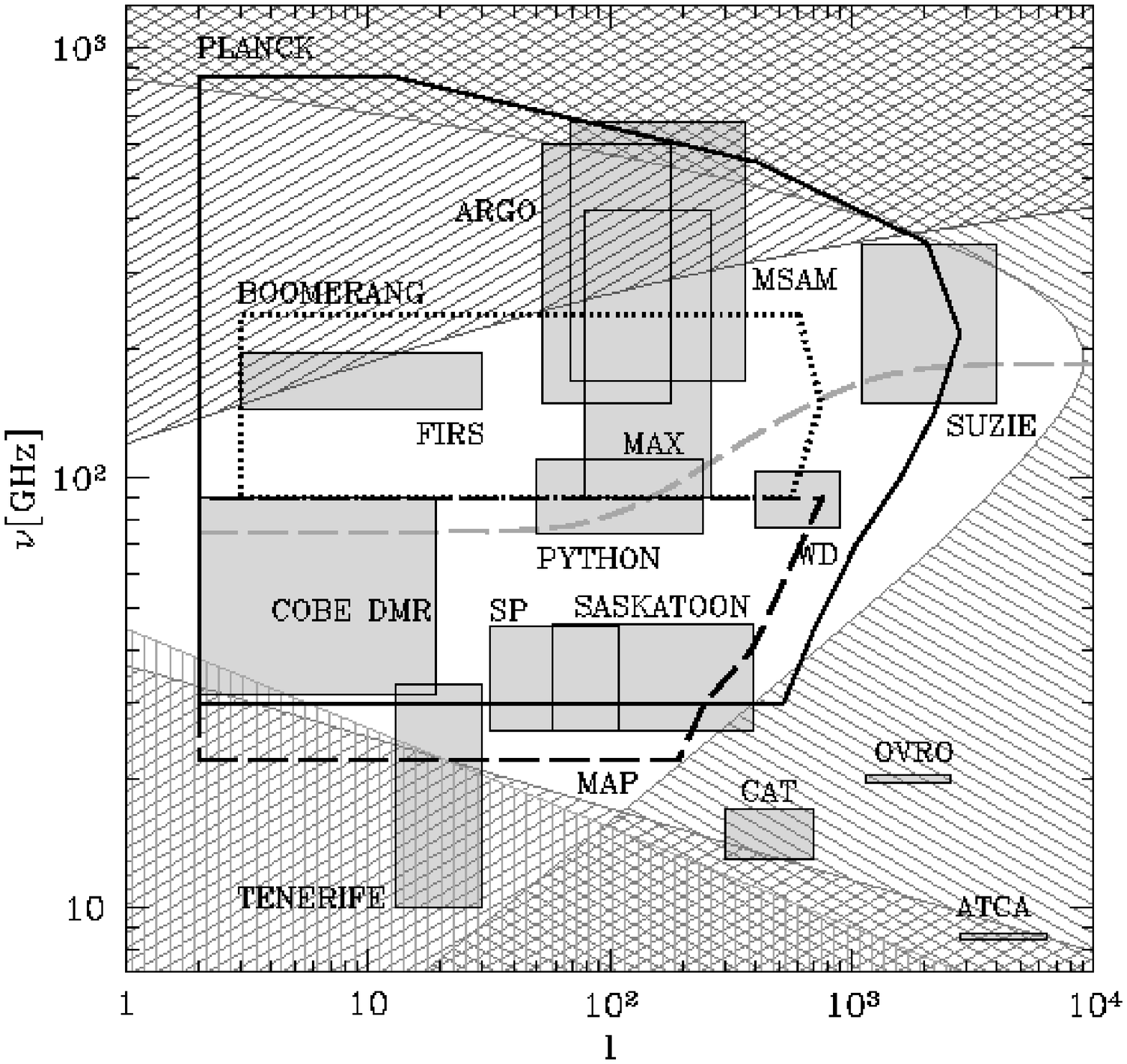}}
\caption[Frequency-multipole range of CMB experiments]
{The boxes indicate the frequency-multipole range probed by 
several CMB experiments. The shaded regions show where
the foregrounds fluctuations are expected to exceed those of the 
cosmological signal in the cleanest 20$\%$ of the sky. They correspond 
to dust (top), free-free emission (lower left, vertically shaded), 
synchrotron(lower left) and point-sources (lower and upper
right). The heavy dashed line shows where the total 
foreground contribution to each multipole is minimal.
Figure kindly provided by M.Tegmark.
.}
\label{tegmark}
\end{figure}

A summary of recently completed, current and future experiments
is given in tables~\ref{exp_ground} (ground-based) and~\ref{exp_balloon} 
(balloon-borne).
A list of the web sites related to these CMB experiments 
(including the satellite missions) is given after the bibliography.
For a more detailed description of these experiments see the reviews
of Halpern \& Scott (1999), Smoot (1997),  Lasenby \etal(1998) and 
de Bernardis \& Masi (1998).
In addition, several experiments will measure the polarization
of the microwave sky (see Staggs \& Gundersen 1999 for a description). 
In Figure~\ref{tegmark}, kindly provided by M.Tegmark, the 
frequency-multipole range covered by various CMB experiments is given.
The shaded regions indicate where the different foregrounds
(see \S~\ref{foregrounds}) are expected to dominate over the 
cosmological signal.

MAP will measure the microwave sky in five different frequencies ranging
$22-90$ GHz with a resolution $\sim 20'$ and a sensitivity
of $35 \mu K$ per $0.3^\circ \times 0.3\deg$ pixel 
during one year of continuous observation.
This sensitivity is expected to be increased to $\sim 20 \mu K$
when combining the three highest frequency channels.
This will allow to map the power spectrum up to $\ell$'s$\sim 800$. 
In Table~\ref{map}, the main characteristics of the MAP satellite are 
given.

\begin{table}[t]
\begin{center}
\caption{MAP instrument description}
\label{map}
\begin{tabular}{|c|c|c|c|c|}
\hline
Frequency & Wavelength & FWHM & No. of & Sensitivity $(\mu K)$ \\
 (GHz)    &    (mm)    &    & channels & $0.3\deg \times 0.3\deg pixel$ \\
\hline
22 & 13.6 & $0.93\deg$ & 4 & 35 \\
30 & 10.0 & $0.68\deg$ & 4 & 35 \\
40 & 7.5  & $0.47\deg$ & 8 & 35 \\
60 & 5.0  & $0.35\deg$ & 8 & 35 \\
90 & 3.3  & $0.21\deg$ & 16 & 35 \\
\hline
\end{tabular}
\end{center}
\end{table}

The Planck satellite is constituted by two different instruments:
the Low Frequency Instrument (LFI) and the High Frequency 
Instrument (HFI).
The LFI will use HEMT technology and will measure the microwave sky
at frequencies 30-100 GHz. The HFI will be based on bolometers and will
cover frequencies from 100-900 GHz. 
Planck will provide multifrequency all-sky maps at a resolution 
$\sim 10'$ and a sensitivity $\frac{\Delta T}{T} \sim 2 \times 10^{-6}$.
The characteristics of the proposed Planck payload are
summarized in table~\ref{planck} (Tauber 1999).

\begin{table}[t]
\begin{center}
\caption{Characteristics of proposed Planck payload}
\label{planck}
{\scriptsize
\begin{tabular}{|c|c|c|c|c|c|c|c|c|}
\hline
Inst.& Freq. & Angular & Detector & Detector  &No. of & 
Bandwidth &Sensitivity$^{\ast}$ & Sensitive to \\
  &(GHz)    &  resolution & technology     & temp.(K) &detectors & 
$(\Delta\nu/\nu)$ &
& linear pol.\\
\hline
LFI & 30  & $33'$ & HEMT & $\sim 20$ & 4 & 0.2 & 1.6 & yes \\
LFI & 44  & $23'$ & HEMT & $\sim 20$ & 6 & 0.2 & 2.4 & yes \\ 
LFI & 70  & $14'$ & HEMT & $\sim 20$ & 12 & 0.2 & 3.6 & yes \\
LFI & 100 & $10'$ & HEMT & $\sim 20$ & 34 & 0.2 & 4.3 & yes \\
\hline
HFI & 100 & $10.7'$ & Bol & 0.1 & 4 & 0.25 & 1.7 & no \\
HFI & 143 & $8.0'$ & Bol & 0.1 & 12 & 0.25 & 2.0 & yes \\
HFI & 217 & $5.5'$ & Bol & 0.1 & 12 & 0.25 & 4.3 & yes \\
HFI & 353 & $5.0'$ & Bol & 0.1 & 6 & 0.25 & 14.4 & no \\
HFI & 545 & $5.0'$ & Bol & 0.1 & 8 & 0.25 & 147.0 & yes \\
HFI & 857 & $5.0'$ & Bol & 0.1 & 6 & 0.25 & 6670 & no \\
\hline
\end{tabular}
\begin{itemize}
\item [$\ast$] Average $\frac{\Delta T}{T}$ per resolution element
(12 months of observation, $1\sigma$, $10^{-6} units)$
\end{itemize}
}
\end{center}
\end{table}

\section{Initial density perturbations}
\label{initial perturbations}
The study of the CMB temperature anisotropies is tightly related
to the initial matter density (scalar) perturbations.
Initial vorticity (rotational) perturbations decay as the universe expands 
and, therefore, are not relevant. 
Initial gravitational wave (tensor) perturbations can 
leave their imprint in the CMB (we will comment about this in
\S\ref{gravitational}).
According to the present theories of galaxy formation, all the 
structures of the universe would have formed from these initial 
density perturbations that eventually would have collapsed 
via gravitational instability.
There are two main scenarios that try to explain the formation of these 
initial seeds: inflation and topological defects.

In the inflationary paradigm (Guth 1982, Linde 1982,1983, 
Albrecht $\&$ Steinhardt 1982) these fluctuations are originated
from quantum fluctuations which are boosted in scale during the exponential 
expansion that characterizes the inflationary epoch
(Hawking 1982, Guth $\&$ Pi 1982, Starobinskii 1982,
Bardeen \etal 1983). 
Inflation can also explain why the universe is so homogenous at large
scale. Regions that appear today causally disconnected, were in causal
contact before the exponential inflation. On the other hand, 
a flat universe arises naturally from inflation, which would explain 
why the value of $\Omega$ is so close to 1 at present.
Standard inflation also predicts that
the primordial density perturbations are a 
realization of a homogeneous and isotropic 
Gaussian random field, which leaves a Gaussian
imprint in the CMB.

An alternative mechanism to induce initial perturbations are 
topological defects, such as cosmic strings, textures and 
monopoles, which may form during symmetry breaking phase transitions in the
early Universe (for a review see Vilenkin \& Shellard 1994).
An important property of the CMB temperature fluctuations induced
by topological defects is their non-Gaussian character. Therefore,
testing the Gaussianity of the CMB would allow to discriminate
between the inflationary and topological defects scenarios.

However, appreciable deviation from non-Gaussianity can also 
arise in non-standard inflationary models (Salopek 1992,
Peebles 1999a,b).
On the other hand, hybrid scenarios that combine inflation and 
topological defects have also been proposed (Jeannerot 1996, 
Linde \etal 1997, Avelino \etal 1998).

Two different classes of initial density fluctuations are distinguished:
{\it adiabatic} and {\it isocurvature}. The adiabatic fluctuations are
characterised by a null fluctuation of specific entropy associated
to each component $\delta\left(\frac{n_b}{n_\gamma}\right)
=\delta\left(\frac{n_x}{n_\gamma}\right) = 0$ at each point, 
which implies the following relation at the initial time:
\begin{equation}
\label{adiabatic}
\delta_\gamma=\frac{4}{3}\delta_b=\frac{4}{3}\delta_X 
\end{equation}
where $\delta_\gamma$, $\delta_b$ and $\delta_X$ 
denote the initial density fluctuations associated to photons, 
baryons and non-baryonic dark matter component, respectively. 
The isocurvature fluctuations  are characterized by a null fluctuation 
of total energy at each point, i.e., $\delta(\rho_\gamma+
\rho_b+\rho_X)=0$, what keeps constant the space curvature.
In addition it is usually assumed that the entropy per baryon
remains constant, i.e. $\delta\left(\frac{n_b}{n_\gamma}\right)=0$,
when there exists a non-baryonic dark matter component.
This leads to the following relation at the initial time:
\begin{equation}
\label{isocurvature}
\delta_\gamma=-\frac{4\rho_X}{3\rho_b+4\rho_\gamma}\delta_X
\end{equation}
The inflationary models favour adiabatic fluctuations (e.g. 
Kolb \& Turner, 1990) with a Harrison-Zel'dovich spectrum (see below)
but isocurvature fluctuations are also possible
(Efstathiou \& Bond 1986, Peebles 1999a,b).

In the case of Gaussian fluctuations, they are fully
characterised by their power spectrum P(k).
This quantity reflects how the amplitude of the fluctuation depends
on the wavenumber $k$ (or equivalently, on the scale):
\begin{equation}
P(k) = \left<|\delta(k)|^2\right> = Ak^n \, ,
\end{equation}
where $\delta(\vec{k})$ are the Fourier components of the density
fluctuation field, 
$A$ is a normalization constant and $n$ is the spectral index.
For the case of adiabatic fluctuations, $n=1$ corresponds to 
the so-called Harrison-Zel'dovich (Harrison 1970, Zel'dovich 1972)
or scale invariant spectrum. 
This scale invariance means that the amplitude of the density 
fluctuations is the same when it enters the horizon scale.

In addition to baryonic matter, non-baryonic {\it dark matter}
is believed to be present in our universe. This idea is strongly
supported by different observations
(e.g. White \etal 1993, Tyson \etal 1998, see
also van den Bergh 1999 and references therein).
Two candidates for this kind of matter have been mainly considered.
Cold dark matter (CDM) is constituted by 
weakly interacting particles whose velocity dispersion is negligible
compared to that of galaxies at the epoch of galaxy formation,
such as WIMP's (weakly interactive massive particles) or
the axion.
On the other hand, hot dark matter (HDM) is constituted by particles
that keep relativistic up to recent times, such as light
neutrinos with non-zero mass.
We must point out that baryon perturbations evolve differently 
from perturbations in cold or hot dark matter.

\section{Sources of temperature anisotropies}

Density perturbations leave their imprint as temperature anisotropies 
in the CMB. Their statistical properties can
be calculated for the inflationary paradigm with an accuracy 
better than $1\%$ (e.g. Seljak \& Zaldarriaga 1996, Hu \etal 1998). 
However, the situation becomes much more complicate for the 
topological defects scenarios due to the non-linear evolution
of the defects and their active role seeding anisotropies in the CMB.
(e.g. Pen \etal 1997).
We will not discuss topological defects any further but
we would like to point out that if such defects exist, they 
would produce anisotropies not only until the epoch of recombination
but also at latter times.
For instance, cosmic strings would generate in the temperature field
a steplike discontinuity along the direction of the string, known as 
the {\it Kaiser-Stebbins} effect (Kaiser \& Stebbins 1984).

In this section, we summarize the main sources that generate 
temperature anisotropies in the CMB (for a review see Hu \etal 1997) .
The anisotropies are usually categorized according to their origin. 
Primary anisotropies are generated until the decoupling time, 
whereas secondary anisotropies are imprinted in the CMB during 
the way of the photons from the last scattering surface (LSS) to us.

\subsection{Primary anisotropies}
\label{primary}
Primary anisotropies in the CMB are comprised of three different 
contributions (e.g. Mart\'\i nez-Gonz\'alez \etal 1990). 
Then, if recombination occurs at the matter-dominated era
and $\Omega \simgt 0.1$:
\begin{equation}
\frac{\Delta T}{T}(\vec{n}) \approx \frac{1}{4}\delta_{\gamma d}+
\frac{1}{3}\phi_d - \vec{n} \vec{v_d}
\end{equation}
where $\vec n$ is the direction given by the line of sight and the
subscript $d$ indicates quantities at decoupling time (using units
with $c=8\pi G \equiv 1$).
The first term represents the anisotropy due to the photon density 
fluctuations at recombination (given approximately by equations 
(\ref{adiabatic}) and (\ref{isocurvature}) 
at decoupling for adiabatic and 
isocurvature perturbations, respectively).
The gravitational redshift of photons climbing out of
potential wells in their way from the LSS to us is 
given by the second term (known as Sachs-Wolfe effect). 
The third term represents the Doppler effect due to the 
peculiar velocities of the last scatterers of the photons.

The combination of these three terms, which are model-dependent,
determines the main features of the angular power spectrum.
The Sachs-Wolfe effect (Sachs \& Wolfe 1967) dominates at scales larger 
than the horizon size at decoupling, 
$\theta \simgt 2\deg\Omega^{1/2}$. 
At these scales, the initial perturbations can not be affected
by causal processes and therefore, the CMB anisotropies 
(at $\ell \simlt 20)$ are
directly related to the initial power spectrum of matter density 
fluctuations. If $P(k) \propto k^n$ then (Bond \& Efstathiou 1987):
\begin{equation}
\cl^{SW}=Q^2_{\rm rms-PS} \frac{4\pi}{5}
\frac{\Gamma\left(\ell+\frac{n-1}{2}\right)
\Gamma\left(\frac{9-n}{2}\right)}{\Gamma\left(\ell+\frac{5-n}{2}\right)
\Gamma\left(\frac{3+n}{2}\right)} \, .
\end{equation}
where $Q^2_{\rm rms-PS}$ is the quadrupole normalization.
In particular, for a Harrison-Zel'dovich spectrum 
$\cl \propto 1/(\ell(\ell+1))$.
This is reflected in the angular power spectrum (per logarithmic scale)
as a {\it plateau} at low $\ell$ (see figure~\ref{datos_exp}).
These large scales are the ones probed by COBE. 
Several independent methods have been used to fit the quadrupole
normalization and spectral index using the COBE data, which have leaded to
consistent results with each other (Hinshaw \etal 1996, G\'orski 1994, 
G\'orski \etal 1996, Wright \etal 1996; see also Bennett \etal 1996 for 
a summary). For instance, G\'orski \etal (1996) obtain
a quadrupole normalization $Q_{\rm rms-PS}=15.3^{+3.7}_{-2.8}\mu K$
and a spectral index $n=1.2\pm 0.3$, which is
compatible with a Harrison-Zel'dovich spectrum. 

The anisotropies generated 
at intermediate angular scales ($0.1\deg \simlt \theta \simlt 2\deg$) 
are directly related to small-scale
processes (inside the horizon) occurring until decoupling time.
In this way, acoustic oscillations of the baryon-photon fluid give 
rise to the so-called {\it Doppler peaks}. 
The position of these peaks is mainly determined by the geometry 
of the universe (e.g. Kamionkowski \etal 1994) due to the fact that the same 
physical scale subtends different angular scales depending on the
curvature. For open universes, the angle is smaller and for 
closed universes is larger than in the flat case. Therefore, the 
Doppler peaks are shifted to smaller scales (larger $\ell$'s) for 
smaller values of $\Omega$ (as can be seen in figure~\ref{datos_exp}).

At small scales, temperature fluctuations are damped due to the
fact that decoupling is not instantaneous, i.e., the LSS has a
finite thickness $\Delta z \sim 100$ (Jones \& Wyse 1985). Thus, 
fluctuations with smaller scales than the thickness of the LSS will 
be reduced by averaging over photons coming from near and far parts 
of the LSS. This corresponds to scales $\theta \simlt 10'\Omega^{1/2}$.

On the other hand, there are processes that decrease the 
matter and radiation density fluctuations at the LSS, therefore 
affecting the temperature anisotropies of the CMB. 
One of these mechanisms is the `Silk damping' of adiabatic 
baryonic perturbations (Silk 1968). Photons diffuse out of
overdense regions, `dragging' the baryons with them. Therefore,
a decrease in the density fluctuations of both 
baryons and radiation is produced. 
Silk damping only operates at small angular scales and
does not affect any kind of present non-baryonic dark matter, 
since this matter is not coupled to radiation.
However, there is a second process that can reduce the fluctuations
of non-interacting, collisionless particles, such as the hypothetic
constituents of dark matter (Bond \& Szalay 1983). 
This mechanism is `free streaming' of collisionless particles 
from high to low density regions. The damping of density fluctuations
by this mechanism depends on the mass and velocity of the particles
involved. Cold particles are not significantly affected 
by this process, since they move too slowly. 
On the other hand, the scale of density fluctuations affected by 
free streaming in the case of hot particles depends on their mass.
For instance, for neutrinos with mass $m_\nu \sim 10 eV$, 
density fluctuations can be strongly damped at scales corresponding to a 
supercluster of galaxies today (Bond \& Szalay 1983).

\subsection{Secondary anisotropies}

Different processes occurring in the way of the photons from the LSS to the 
observer can generate secondary anisotropies in the CMB.
Thus, they provide information about the evolution of the 
universe after decoupling.
They can be categorized in gravitational and rescattering effects
(for a more detailed description see Hu 1996, Hu \etal 1997).

\subsubsection{Gravitational effects}
\label{gravitational}

Gravitation can induce secondary anisotropies in the CMB temperature
field in different ways.
One of such ways is the integrated Sachs-Wolfe (ISW) effect. 
When a photon falls in and climb out of a potential well, constant
in time, the net change in the energy of the photon is zero. However, 
if the depth of the potential well varies as the photon crosses, the
blueshift from falling in and the redshift from climbing out
no longer cancel. The magnitude of the ISW is given by an integral
along the photon's path (Mart\'\i nez-Gonz\'alez \etal 1990):
\begin{equation}
\frac{\Delta T}{T} = \int 
\frac{\partial\phi}{\partial t}(\vec{r},t)dt \, .
\end{equation}
On the other hand, gravity can also deflect the trajectory of a photon
but without modifying its energy (gravitational lensing).
The different cases that generate secondary anisotropies 
via these effects can be summarized as follows:

\begin{itemize}
\item[(i)] In typical models, the epoch of matter-radiation equality
occurs before recombination but not long before. Thus the photon 
contribution
to the density of the universe is not completely negligible at last 
scattering and shortly thereafter. 
The decay in the potential shortly after last scattering 
gives rise to the {\it Early} ISW effect. This effect contributes
at scales just larger than the first acoustic peak.
\item[(ii)] In open or $\Lambda$ models, the potential decays at late
times, typically at redshifts $z \simlt \Omega^{-1}$. 
This produces the so-called {\it Late} ISW effect, that also shows
up on large angular scales.
\item[(iii)] At late times, evolving non-linear structures cause the 
potential to vary with time. This kind of ISW effect 
is usually called the {\it Rees-Sciama} 
effect (Rees \& Sciama 1968, Sanz \etal 1996).
In standard CDM models, its contribution to the radiation power spectrum
seems to be negligible except at very small angular scales. Thus, it is 
not likely to be detected by the next generation of satellites.
\item[(iv)]
The ISW effect changes the energy of the photons but not their 
direction of motion. However, gravity can also produce the opposite
effect via {\it gravitational lensing}: the trajectory of the photons 
is deflected but their energy is left unchanged. 
This effect slightly distorts the image of the LSS, producing
a smearing of the angular power spectrum, 
with power from the peaks being moved into the valleys
(Mart\'\i nez-Gonz\'alez \etal 1997).
Although this effect is typically weak (a few percent change in the power
spectrum), it could be detectable by some future CMB experiments.
\item[(v)]
Other possible sources of secondary anisotropies are gravitational 
waves.
The magnitude of the ISW effect is given in this case by an integral
of the time derivative of the invariant metric perturbation along the 
photon's path (Sachs \& Wolfe 1967).
The gravitational waves would affect the radiation power spectrum at
scales larger than the horizon at recombination (Crittenden \etal
1994).

\end{itemize}

\subsubsection{Scattering effects from reionization}
\label{reionizacion}
Reionization of the universe after recombination
produces free electrons that rescatter off the photons of the
microwave background. Therefore, primary anisotropies are 
washed out and new secondary ones appear.
If the universe becomes globally reionized at high redshift,
primary anisotropies can be dramatically suppressed.
On the other hand, local reionization
also produces characteristic features in the CMB.
For a recent review on reionization and its effects on the CMB see
Haiman \& Knox (1999).

\begin{itemize}
\item[(i)] If the universe becomes globally reionized at a given redshift
$z_r$, a certain fraction of 
the CMB photons will be rescattered by the free electrons.
Therefore, a photon coming toward us from a particular direction, 
has not necessarily been originated from that direction.
Thus, each location of the sky contains contributions from photons 
coming from different regions of the LSS, producing a damping of the 
fluctuations. 
The scales affected by this smearing are those smaller than the
horizon size at the redshift $z_r$ of the rescattering epoch.
On the other hand, the fraction of CMB photons that are never
rescattered is $e^{-\tau}$, where 
$\tau \equiv {\sigma}_T\int dt\,n_e $ 
($n_e$ is the electron density 
and ${\sigma}_T$ is the Thomson cross-section) is the optical depth
(see for instance Tegmark \& Silk 1995).
\item[(ii)] Another source of secondary anisotropies in reionized
universes is the so-called {\it Vishniac} effect (Ostriker \& Vishniac
1986, Vishniac 1987). This is a 
second-order effect due to coupling 
between the bulk flow of the electrons and their density 
fluctuations that generates new anisotropies at very small 
angular scales.
\item[(iii)] 
Inverse Compton scattering of microwave photons by hot 
electrons in the intracluster gas of a cluster of galaxies 
produces spectral distortions in the 
blackbody spectrum of the CMB, known as the {\it thermal 
Sunyaev-Zel'dovich} (SZ) effect (Sunyaev \& Zel'dovich 1970,1972).
In addition, the peculiar velocities of clusters also produces secondary 
anisotropies in the CMB via the Doppler effect, known as the {\it kinetic 
SZ} effect (Sunyaev \& Zel'dovich 1980).
For a recent review on the SZ effect, see Birkinshaw 1999.

The thermal SZ has a characteristic frequency dependence.
CMB photons that interact with electrons in the intracluster gas
gain energy via inverse Compton scattering. This generates a
decrease in the number 
of photons at frequencies lower than $\sim$ 217GHz and an 
increment at higher frequencies. 
The change of spectral intensity is given by:
\begin{equation}
\Delta I = \frac{2(kT_o)^3}{(hc)^2}\frac{x^4e^x}{(e^x-1)^2} y \left[ 
xcoth\frac{x}{2}-4\right] \quad , \quad x=\frac{h\nu}{kT_o}
\end{equation}
where 
$y \equiv \frac{k{\sigma}_T}{m_e}\int dl\ T_e n_e$
is the Comptonization parameter and is a function
of the electron density $n_e$ and temperature $T_e$.
This spectral dependence will help to separate
the thermal SZ effect from the intrinsic cosmological 
signal in multifrequency observations.

On the other hand, the temperature fluctuation originated by the
kinetic SZ effect is given by:
\begin{equation}
\frac{\Delta T}{T}=-\tau\frac{v_r}{c}
\end{equation}
where $v_r$ is the radial velocity of the cluster and $\tau$
is the optical depth.

The scales affected by these effects are those of the hot gas of 
cluster of galaxies, i.e., below a few arcmin.
The amplitude of the thermal SZ effect is expected to be  
at the level of $\sim 10^{-6}$ at these scales 
whereas that of the kinetic SZ is $\sim 10$ times smaller.
In addition, the kinetic SZ has the same spectral dependence
that the CMB, because it is just a Doppler shift.
This will make very difficult to separate this effect from
the cosmological signal.
On the other hand the angular 
power spectrum of the secondary anisotropies generated by the SZ 
effects is, approximately, that of a Gaussian noise, i.e., 
$\cl$ constant with $\ell$.

Future space missions will observe a large number of clusters
via the SZ effect. In particular, it is 
expected that the Planck mission will detect tens of thousands of 
clusters, although the exact number depends on the cosmological model
(Aghanim \etal 1997). These data, combined with X-ray observations, 
will lead to an independent measurement of the Hubble constant
(Cavaliere \etal 1977).
In addition, all this information will provide an interesting
tool to study different properties of the clusters, such as
the cluster peculiar velocities (Haehnelt \& Tegmark 1996).

\item[(iv)] 
It has been recently discussed that inhomogeneous
reionization due to early formed stars or quasars could produce
second order CMB fluctuations at subdegree scales
through the Doppler effect 
(Aghanim \etal 1996, Gruzinov \& Hu 1998, Knox \etal 1998, Peebles \&
Juszkiewicz 1998). However, calculations of their amplitudes are still 
highly uncertain and it is not clear whether they
would significantly affect parameter estimation from the 
angular power spectrum of the CMB.

\end{itemize}

\section{Foreground emissions}
\label{foregrounds}
The microwave sky contains not only contribution from the CMB but also
from several foreground components. In particular, the main
foreground components are Galactic dust, free-free (bremsstrahlung)
and synchrotron emission together with extragalactic point sources
(for reviews see Davies 1999, Bouchet \& Gispert 1999, Tegmark \etal 1999).
In addition, a fourth Galactic foreground produced by spinning
dust grains (Draine \& Lazarian 1998) could also be present, being
detectable at frequencies $\sim10-100$ GHz.
In order to obtain all the valuable information contained on the CMB, 
it is necessary to separate the different contributions of the 
foregrounds from the cosmological signal. In addition, the 
foregrounds contain themselves very valuable information on astrophysical
phenomena. Therefore, studying their contribution 
to the microwave sky becomes of great interest (see De Zotti 1999).
In figure~\ref{tegmark}, the ranges of frequencies and multipoles
where the different foregrounds dominate are shown for the $20\%$
cleanest region of the sky.
It can be seen that at low $\ell$'s Galactic contributions are more
important, whereas point sources mainly affect at 
high $\ell$'s. On the other hand, the fluctuations produced
by foreground emission present, in general, a non-Gaussian behaviour.

\subsection{Synchrotron}

Synchrotron emission is produced by relativistic electrons that are
accelerated in magnetic fields (for reviews see Davies \& Wilkinson 1998, 
Smoot 1999).
Therefore, it depends on the energy spectrum of the electrons and
on the intensity of the magnetic field (see Longair 1994).
This component dominates the Galactic emission
at low frequencies $\nu \simlt 20$ GHz.
The brightness temperature of synchrotron emission (or, equivalently, 
their intensity)
is usually written in terms of a power law,
$T_b \propto \nu^{-\beta_{\mathrm{syn}}}$.
However, the synchrotron spectral index $\beta_{\mathrm{syn}}$ is expected
to vary with frequency and position (Lawson \etal 1987). 

The low frequency surveys by Haslam \etal
(1982) at $408$MHz (the only all-sky map available at these
frequencies) and by Reich \& Reich (1986) at $1420$MHz have been
normally used to estimate the amplitude of synchrotron emission 
at higher CMB frequencies by extrapolation.
However, both of these maps are affected
by significant uncertainties associated to 
the zero level, the gain stability and scanning errors. 
On the other hand, the spatial information is
limited by their finite resolution,
$0.85\deg$ and $0.6\deg$ for the Haslam and Reich \& Reich maps,
respectively. 

For instance, a recent determination of $\beta_{\mathrm{syn}}$ in the
range 1-10 GHz has been given by Platania \etal (1998).
Combining the Haslam and Reich \& Reich
surveys with their own data obtained at the White Mountain, California 
(Smoot \etal 1985) with an angular resolution of
$18\deg$,
they find a mean spectral index of 
$\beta_{\mathrm{syn}}=2.76\pm 0.11$ for a celestial region 
at declination $36\deg$. Their data also
suggest a steepening of the synchrotron spectrum toward higher 
frequencies.
These results are consistent with previous estimations 
derived from different authors (see Platania \etal 1998 
and references therein)

The angular power spectrum of the synchrotron emission is not
well known. 
It has been suggested that the Galactic power spectrum behaves
as $\ell^{-3}$ as one approaches smaller angular scales 
(Tegmark \& Efstathiou 1996).
However, Lasenby (1997) has estimated the power spectrum of synchrotron
for the high-latitude regions observed by the Tenerife 
experiment from the Haslam and Reich \& Reich maps 
finding an angular power spectrum 
slightly flatter than the $\ell^{-3}$ law. 
 
\subsection{Free-free}

The free-free emission is the thermal bremsstrahlung from hot 
($ T \simgt 10^4$K) electrons when accelerated by ions in the 
insterstellar gas (for reviews see Bartlett \& Amram 1998,
Smoot 1998).
Free-free emission is the less well known Galactic foreground. 
This is due to the fact that it only dominates over a small range of
frequencies ($ \sim 25-75$ GHz), where the total Galactic emission is minimal. 
Therefore, it can not be traced by observing at higher or lower frequencies,
unlike dust and synchrotron emission.
However, diffuse Galactic $H_\alpha$ is thought to be a good tracer 
of free-free emission, since both are emitted by the same ionized medium
(e.g. McCullough \etal 1999). 
The combination of WHAM (Haffner \etal 1998) observations 
with the southern celestial hemisphere $H_\alpha$ survey 
(McCullough \etal 1999) will allow to create in
the near future a spatial template for free-free emission based 
on $H\alpha$ observations.
At present, the only all-sky maps at frequencies of interest to
study free-free are those of COBE.
Bennett \etal (1992,1994) used a combination of the COBE DMR maps
to get a free-free map at 53 GHz assuming a spectral index of 
$\beta_{\mathrm{ff}}=2.15$ ($T_b \propto \nu^{-\beta_{\mathrm{ff}}}$).
Using this technique, they found an amplitude for the 
free-free emission of $\Delta T = (10 \pm 4) \csc(|b|)~\mu K $ 
for $|b|>15\deg$.

On the other hand, several authors (Kogut \etal 1996a, Leitch \etal 1997,
de Oliveira-Costa \etal 1997) have found
an anomalous component of Galactic emission at 15-40 GHz
that correlates with the $100 \mu m$ IRAS and DIRBE maps.
Kogut \etal identifies this emission as a free-free component that correlates 
with the dust emission. However, it has been noted (e.g. Smoot 1998)
that there are inconsistencies between 
the estimated level of free-free from $H\alpha$
and that implied from the correlation between free-free 
and dust emissions.
A possible explanation for these inconsistencies is given by
Draine \& Lazarian (1998). They propose that emission
from rotating dust grain could account, at least partially, for
this anomalous component. This emission would be detectable
at 10-100 GHz and would have a similar spectral dependence to
free-free around $\nu=25$ GHz.
Recently, de Oliveira-Costa \etal (1999) have found the DIRBE-correlated
Galactic emission of the Tenerife experiment to be better explained
by dust rotating grains than by free-free emission.
However, an independent analysis of the same data that is currently 
being performed does not favour either of the possibilities
(Jones, private communication). Future CMB experiments should be 
able to show what this anomalous component actually is.

Regarding the power spectrum of free-free emission, 
there are still uncertainties about its shape.
Kogut \etal (1996a) found that the power 
spectrum of the dust-correlated free-free component
was proportional to $\ell^{-3}$. 
On the other hand, Veeraraghavan and Davies (1997) used $H_\alpha$ maps 
of the North Celestial Pole made by Gaustad \etal (1996)
to estimate the free-free emission on scales of $10'$ to a few degrees.
They found a best fit of 
$C_\ell^{\mathrm{ff}} \propto \ell^{-2.27\pm0.07}$, what it is significantly
flatter than the $\ell^{-3}$ law. However, 
their normalization is also considerably lower than
that derived from Kogut \etal resulting in
a lower signal at all scales of interest.

\subsection{Dust emission}

Dust grains in our galaxy are heated by the interstellar radiation
field, absorbing UV and optical photons and re-emitting the 
energy in the far infrared. The observed dust emission is the 
sum over the emission from each dust grain 
along the line of sight.
This foreground dominates the Galactic emission at $\nu\simgt 90$ GHz.

Dust emission can be modelled by
a modified blackbody emissivity law
$I(\nu) \propto B_\nu(T_d) \nu^\alpha$ 
with $\alpha \simeq 2$ (Draine \& Lee 1984). 
Wright \etal(1991) and Reach \etal(1995) found that two dust components were
necessary to fit the Galactic dust emission, including a cold
component with $T \sim 7 K$.
If this cold dust component exists, it could dominate the dust
emission at low frequencies. However,
the uncertainty about its existence is still very high. 
Kogut \etal (1996b) found that the high-latitude 
dust emission ($|b| > 20\deg$)
of the DMR-DIRBE maps is well fitted by a single dust component with 
temperature $T=18^{+3}_{-7}$ and an emissivity index 
$\alpha=1.9^{+3.0}_{-0.5}$.

Schlegel \etal (1998) combined the IRAS and DIRBE maps to construct
an all-sky dust map at 100 $\mu m$ with an angular resolution of $\sim 6'$.
They also provided a map of dust temperature $T_d$ by adopting
a modified blackbody emissivity law with $\alpha=2$. The dust
temperature varies from $17$ to $21K$.
This map can be used as a dust template to extrapolate to the range of
frequencies probed by future CMB experiments.

Gautier \etal (1992) estimated the dust power spectrum
from the IRAS $100\mu m$ map, finding
$\cl^{dust} \propto \ell^{-3}$ at scales $8\deg-4'$.
This result has been confirmed by
Wright (1998) using the DIRBE maps, who
found $\cl \propto \ell^{-3}$ for $2 < \ell <300$.

\subsection{Extragalactic point sources}

Emission from extragalactic point sources must be taken into account
for the future high-resolution CMB experiments.
Different source populations dominate above and below $\nu \sim 300$ GHz.
At lower frequencies, radio sources give the main contribution
whereas at higher frequencies, far-IR sources dominate.
These populations mainly consist of
compact AGN, blazars and radio loud QSOs in the radio and 
of inactive spirals galaxies in the far-IR.
The number counts and spectral dependence of these populations 
are subject to many uncertainties due to the lack of surveys
in the frequency range explored by CMB experiments. 
Detailed studies of the contribution of point sources
at Planck frequencies have been performed by several authors 
(Toffolatti \etal 1998,1999, Bouchet \& Gispert 1999,
Guiderdoni 1999, Gawiser \& Smoot 1997, Sokasian \etal 1998).
According to the previous authors,
it is expected that Planck will detect from several hundred to many
thousand of sources as $5\sigma$ peaks at each frequency channel.
Once the resolved point sources have been removed, we are left with
a background due to the unresolved point sources.
Following Bouchet \& Gispert the confusion limit due to the
Far Infrared sources background can be approximately treated as another 
template to be extracted from the data. Using the results of Guiderdoni \etal
(1997,1998), the previous authors find the spectral behaviour of 
the infrared background to be approximately modelled as:
\begin{eqnarray}
\ell\cl^{1/2} & \simeq & \frac{7.1 \, 10^{-9}}{e^{x/2.53}-1}
\left(1-\frac{0.16}{x^4} \right) \frac{\sinh^2x}{x^{0.3}}\ell\, [K] \, ,
\nu > 100{\rm GHz} , \nonumber \\
\ell\cl^{1/2} & \simeq & 6.3\, 10^{-9} \left(0.8 -2.5x +3.38x^2 \right)
\frac{\sinh^2x}{x^4}\ell\,[K] \,, \nu < 100{\rm GHz}.
\label{back1}
\end{eqnarray}
where $x=h\nu/2kT_0=\nu/(113.6{\rm GHz})$.
This estimation assumes that point sources have been removed by a 
simple thresholding technique at the 5$\sigma$ level,
but it is expected that their 
contribution can be subtracted down to lower fluxes.
Thus the level of the unresolved point sources background 
given by eq.(\ref{back1}) can be seen as an upper limit to the 
contribution of the far IR population.

On the other hand, 
the previous expression does not take into account the population of 
point sources that dominate in
low frequency and would therefore underestimate the contribution of 
point sources at the range of frequencies probed by the LFI and MAP. 
Assuming that point sources have been subtracted down to 
100mJy, Toffolatti \etal obtain
\begin{equation}
\ell\cl^{1/2} \simeq \frac{ \sinh^2\left(\frac{\nu}{113.6}\right)}
{(\nu/1.5)^{(4.75-0.185\log(\nu/1.5))}} \ell \, [K]
\label{back2}
\end{equation}
for the background due to the population dominating at low 
frequencies.
The previous authors also predict that
fluctuations from point sources will be well
below the expected amplitude of the CMB fluctuations in the 
frequency range $50-200$GHz 
on all angular scales covered by the Planck Mission.

Regarding the power spectrum of point sources, 
this follows that of white noise, i.e.,
$\cl =$constant for all scales, since 
point sources are, in a first approximation, randomly
distributed in the sky. Thus, confusion from point sources 
mainly affects small angular scales (high $\ell$'s) as can be seen
in figure~\ref{tegmark}.
According to Toffolatti \etal, the
fluctuations due to clustering are generally small in comparison with the
Poissonian term but the relative importance of clustering increases
if sources are subtracted down to faint flux limits.

\section{Beyond the power spectrum}

It has been already pointed out the importance of the angular power spectrum
of the CMB, that completely describes the field in the case of Gaussian
fluctuations as predicted by standard inflation.
However, topological defects as well as certain inflationary models
(e.g. Peebles 1999a,b)
offer alternative scenarios of structure formation that give
rise to non-Gaussianity in the CMB.
Moreover, secondary anisotropies, such as the SZ effect, 
foreground contamination and systematics can well imprint 
a non-Gaussian signal in the CMB temperature fluctuations.
Therefore, it becomes apparent the need of looking for non-Gaussianity
in the CMB.
Since the angular power spectrum only gives information about the
2-point correlation function, the use of  
statistics which introduce higher-order information from the random
temperature field is necessary to test Gaussianity.
In addition, these statistics can be used as a consistency check
of the cosmological parameters determined from the angular power spectrum.
Obtaining the set of $\cl$'s
from future experiments, such as MAP or Planck, is not a trivial matter.
Analysis of such complex and large data set
presents a challenge for the existing and anticipated computers
(for a discussion see Borrill 1999). 
Therefore, alternative statistics, computed from the data in a
manner completely independent 
of the power spectrum, provide a useful check on the power spectrum
computations, even in the case of underlying Gaussian fluctuations.

A large number of estimators to test non-Gaussianity in the CMB has
been proposed. Most of them can be (semi)analytically calculated for 
Gaussian random fields, allowing a straightforward comparison of their
expected value with the results computed from the data. 
Simulations are also extensively used to test the power
of the different estimators. 
In this section we review some popular and novel estimators 
that have been suggested to test Gaussianity in the CMB:

\begin{itemize}
\item[(i)] The simplest estimators are those directly related to the 1-point
distribution function such as the {\it skewness} $S_3$ and 
{\it kurtosis} $K_4$ (Luo \& Schramm 1993a) of the temperature fluctuations
field:
\begin{equation}
S_3=\frac{\mu_3}{\sigma^3} \quad , K_4=\frac{\mu_4}{\sigma^4}-3
\end{equation}
where $\sigma$, $\mu_3$, $\mu_4$ are the dispersion, third and fourth 
central moments of the distribution, respectively. Both quantities
are zero for a Gaussian distribution.

Alternatively, the kurtosis of the gradient temperature map
has also been considered by Moessner \etal (1994)
to study the signal imprint by the Kaiser-Stebbins effect.

\item[(ii)] Another usual quantities to characterize non-Gaussianity 
are those based on the n-point correlation functions. 
In particular, the {\it 3-point correlation function} 
(Luo \& Schramm 1993b, Kogut \etal 1996c) is given by:
\begin{equation}
C_3(\theta_1,\theta_2,\theta_3)=\langle T(\mathbf{n_1})T(\mathbf{n_2})
T(\mathbf{n_3})\rangle \, ,
\end{equation}
where $\mathbf{n_1}\cdot\mathbf{n_2}=\cos\theta_1$,
$\mathbf{n_2}\cdot\mathbf{n_3}=\cos\theta_2$ and 
$\mathbf{n_3}\cdot\mathbf{n_1}=\cos\theta_3$.
Equivalently, the {\it bispectrum} can be used to test Gaussianity
(Luo 1994, Heavens 1998, Ferreira \etal 1998, Spergel \& Goldberg 1998a,b).
This quantity
plays the same role with respect to the 3-point correlation 
function than the angular power spectrum with respect to the 2-point 
correlation function.
The bispectrum is defined as:
\begin{equation}
B(\ell_1\ell_2\ell_3,m_1,m_2,m_3) \equiv \langle a_{\ell_1 m1} a_{\ell_2 m2}
a_{\ell_3 m3} \rangle
\end{equation}
Both, the 3-point correlation function and the bispectrum, are zero
for a Gaussian distribution.

\item[(iii)] {\it Minkowski functionals} (Minkowski 1903) 
have also received extensive attention in the
literature (Gott \etal 1990, Torres \etal 1995, Kogut \etal 1996c, 
Schmalzing \& G\'orski 1997, Winitzki \& Kosowsky 1997,
Novikov \etal 1998).
These quantities are local properties of the excursion
sets above a threshold, translationally and rotationally
invariant and additive. Therefore, they can be used for maps with
incomplete or patchy sky coverage. For a 2-dimensional map, there are
three Minkowski functionals: mean fractional area $<a>$ of excursion sets
enclosed by the isotemperature contours,
mean contour length $<s>$ per unit area and mean genus $<g>$ per unit area.
The genus is a purely topological quantity that can be estimated as 
the total number of isolated high-temperature regions minus the 
number of holes in them.
For a homogeneous and isotropic Gaussian field, the Minkowski functionals 
are given by
\begin{eqnarray}
<a> & = & \frac{1}{2}{\rm erfc}\left(\frac{\nu}{\sqrt2} 
\right) \, , \nonumber \\ 
<s> & = & \frac{1}{2\theta_c} \exp^{-\nu^2/2} \, , \nonumber \\
<g> & = & \frac{1}{(2\pi)^{3/2}\theta_c^2}\nu \exp^{-\nu^2/2} \, ,
\end{eqnarray}
where $\nu$ is a threshold defined in units of the field 
dispersion, erfc is the complementary error function and 
$\theta_c=\left(-C(0)/C''(0) \right)^{1/2}$ 
is the coherence angle 
that depends only on the 2-point correlation function.

\item[(iv)] In addition to Minkowski functionals, 
several properties of excursions sets and maxima have been studied.
They include the number and mean size (Sazhin 1985, 
Zabotin \& Nasel'skii 1985, Bond \& Efstathiou 1987,
Vittorio \& Juszkiewicz 1987, Mart\'\i nez-Gonz\'alez \& Sanz 1989), 
eccentricity and Gaussian curvature (Barreiro \etal 1997, 
Mart\'\i nez-Gonz\'alez \etal 1999, Barreiro \etal 1999)
and the probability density function of the hottest spot
(Coles 1988).
All the considered quantities can be (semi)analytically calculated for
a Gaussian field. The mean area and number of excursion sets have
also been calculated for some non-Gaussian fields derived from
the Gaussian one (Coles \& Barrow 1987).
In particular the expected number of excursion sets $<N_s>$ over the whole 
sphere and their expected area $<A>$ for
a homogeneous and isotropic Gaussian field can be estimated in
the form (Vanmarcke 1983):
\begin{eqnarray}
<N_s> & = & \frac{2}{\pi\theta_c^2}\frac{e^{-\nu^2}}
{{\rm erfc}(\nu/\sqrt{2})}
\nonumber \, ,\\
<A> & = & \left(\pi \theta_c \exp(\nu^2/2) {\rm erfc (\nu/\sqrt{2})} 
\right)^2 \, .
\end{eqnarray}

Following the same direction, properties of clustering of maxima 
(Novikov \& J$\o$rgensen 1996), correlation of maxima
(Bond \& Efstathiou 1987, Kogut \etal 1995, Heavens \& Sheth 1999) 
and correlation of excursion sets 
(Barreiro \etal 1998) have also been studied.

\item[(v)] Multifractal (Pompilio \etal 1995), partition function
(Diego \etal 1999, Mart\'\i nez-Gonz\'alez \etal 1999) and roughness surface 
(Mollerach \etal 1999) based analysis
are additional possibilities in the sought of non-Gaussianity.
These methods provide an alternative way of studying the structure
of CMB maps appearing at different scales.

\item[(vi)] So far, most of the described estimators are defined
in real space. An interesting possibility is to use
estimators based on Fourier statistics. 
A non-Gaussian signal can show up only in a particular scale, and the use
of Fourier space allows to study each scale separately. On the contrary,
spatial information is mixed up. An example of this kind of
estimators is the so-called {\it non-Gaussian
spectra} introduced by Ferreira \& Magueijo (1997), which provides
a way to characterize generic non-Gaussian fields. These quantities
are extracted out of the angular distribution of the Fourier
transform of the temperature anisotropies and take a simple form in 
the case of a Gaussian field. Another possibility is using a double 
Fourier transform, as proposed by Lewin \etal (1999).

\item[(vii)] A promising novel possibility is the use of statistics based
on wavelet techniques. Wavelets coefficients provide simultaneous
information on the position and scale of the temperature field.
Thus, they become
a very useful tool to detect non-Gaussian signals that show up in
a particular scale and in a given region of the sky.
The wavelet coefficients of a Gaussian random field are also Gaussian
distributed at each given scale. Thus, 
the skewness and kurtosis of those coefficients
can be used to detect non-Gaussianity (Pando \etal 1998,
Hobson \etal 1998b).
Statistics based on higher order moments of temperature maps (or 
equivalently on cumulants) have been studied by Ferreira \etal (1997) 
defined in wavelet and Fourier spaces.
Another possibility is the study of the scale-scale correlation 
of the temperature field through wavelet coefficients 
(Pando \etal 1998).

\end{itemize}

The power of a given estimator in detecting non-Gaussianity will strongly
depend on the considered non-Gaussian field.
A particular estimator can perform very well in detecting some kind of
non-Gaussian signal and fail to detect a different one.
They have been usually applied to different cases and
a comparative study becomes necessary before general 
conclusions can be established.
A detailed description of the different tests performed for each of
the proposed non-Gaussian estimators is out
of the scope of this review.
However, we will refer here to the tests of Gaussianity 
that have been applied to the 
COBE data. Their different performance can give us some hints about
the power of the corresponding method.
Torres (1994) used the number of hot spots and their genus to study
the 1-yr COBE data. He found the data to be consistent with being
derived from a parent Gaussian distribution.
Kogut \etal (1995,1996c) tested their Gaussianity through the genus, the
3-point correlation function and the extrema correlation. The result
is again that the data are compatible with a Gaussian distribution.
Moreover, comparison with non-Gaussian toy models show the Gaussian
distribution to be the most probable one of all the considered
cases. They also find the extrema correlation to be a better discriminator
than the genus and 3-point correlation function.
Schmalzing \& Gorski (1997), Heavens (1998) and Diego \etal (1999)
also found the COBE data to be consistent with a Gaussian field
using the Minkowski functionals, the bispectrum and a partition
function analysis, respectively.
However, detection of non-Gaussianity in the COBE data has been 
reported by two groups, although the origin of this non-Gaussian
signal may well not be cosmological. 
Ferreira \etal (1998) studied the 
distribution of an estimator for the normalized bispectrum, finding that
Gaussianity is ruled out at the confidence level $>98\%$. 
However, Banday \etal (1999) showed in a recent work that the 
non-Gaussian signal detected by Ferreira \etal (1998) is likely due 
to a systematic effect.
On the other hand, Pando \etal (1998) performed a discrete wavelet analysis
based on scale-scale correlation and find a significant 
non-Gaussian signal that rules out Gaussianity at the $99\%$ level, 
although the same authors do not find significant deviation from
Gaussianity using the skewness and kurtosis of wavelet coefficients
at each scale. However, the same analysis has been independently performed by
Mukherjee \etal (1999), who are unable to reproduce the numerical results 
obtained by Pando \etal and do not find strong evidence for non-Gaussianity 
in the scales probed by COBE. The previous authors conclude that 
wavelet analysis only rule out Gaussinity at the $76\%$ level.
For a discussion of the statistical significance of these detections
of non-Gaussianity see Bromley \& Tegmark (1999).

Finally we would like to go back to the use of alternative estimators
as an independent consistency check of the computed results from the 
power spectrum. 
Several authors have used alternative estimators to the 2-point
correlation function to fit the amplitude and spectral index
of the power spectrum of the COBE data.
In this direction, Torres \etal (1995) performed an analysis based on
the genus and number of spots, Diego \etal (1999) used a partition 
function technique and Mollerach \etal (1999) studied the roughness
of the temperature surface. 
All these works report consistent results with the ones found by 
more standard methods.
Park \etal (1998) studied the 2-point correlation function and genus
of Gaussian simulations of MAP data. They found the genus to be a good
independent test of the cosmological parameters computed from the power
spectrum.
Barreiro \etal (1997) pointed out the interest of studying topological 
quantities of CMB maps to discriminate between different $\Omega$ values 
and Wandelt \etal (1998) developed tools for their study on 
high-resolution CMB simulations.

\section{Image reconstruction methods}

Future CMB data will provide an extremely valuable information.
However, these maps 
will contain not only the cosmological signal but also the contribution
from the different foregrounds (see $\S$~\ref{foregrounds}) together with 
instrumental noise. Therefore, before extracting all the information
coded in the CMB, `cleaning' of the maps and a separation of
the microwave sky components must be performed.

Many methods have been proposed in Astronomy and other fields 
to reconstruct a signal from noisy data. In the present section we will 
outline the basis of three of these methods as applied to a general image
and comment on their performance when applied to CMB simulations.
For a comparison of different methods as applied to CMB,
see Tegmark (1997) and Jones (1998).

\subsection{Wiener Filter}

Consider a set of N measured data $\mathbf{d}=(d_1,d_2,...,d_N)$ that we 
use to estimate the underlying signal or image 
$\mathbf{s}=(s_1,s_2,...,s_M)$.
We will consider the case in which the data vector $\mathbf{d}$ can 
be written as a linear convolution of the signal:
\begin{equation}
\mathbf{d}=\mathbf{Rs}+\mathbf{n} \, ,
\end{equation}
where $\mathbf{R}$ is some known $N\times M$ matrix and $\mathbf{n}$ is a 
random noise vector. $\mathbf{R}$ usually represents the antenna response 
of the instrument to the underlying field, although it can also include 
a more complicated relationship between the measured data and the signal.

In order to apply Wiener Filter, some assumptions are usually made.
The noise and signal are taken to have zero mean and we assume knowledge
of the covariance matrices:
\begin{eqnarray}
\mathbf{S} & = & <\mathbf{s}\mathbf{s}^t> \, ,\nonumber \\
\mathbf{N} & = & <\mathbf{n}\mathbf{n}^t> \, .
\end{eqnarray}
In addition, signal and noise are taken to be uncorrelated:
\begin{equation}
<\mathbf{n}\mathbf{s}^t>=0 \, .
\end{equation}
Given an estimator of the signal $\mathbf{\hat{s}}$, the
reconstruction error is given by:
\begin{equation}
\bepsilon =\mathbf{s}-\mathbf{\hat{s}} \, .
\end{equation}
Wiener filter (Wiener 1949) is defined as the linear filter $\mathbf{W}$, 
i.e. $\mathbf{\hat{s}}=\mathbf{Wd}$, that minimises the variance of the 
reconstruction error $<|\bepsilon|^2>$.
Carrying out the minimisation, $\mathbf{W}$ is found to be 
(e.g. Rybicki \& Press 1992)  :
\begin{equation}
\mathbf{W}=\mathbf{SR}^t\left(\mathbf{RSR}^t+\mathbf{N}\right)^{-1} \, .
\end{equation}

If the signal and noise are Gaussian random variables, this filter can also
be obtained as the Bayesian estimator of the signal 
(e.g. Bunn \etal 1996).
Derivations of Wiener filter in Fourier and harmonic space 
can be found for instance in Press \etal (1994) and Bunn \etal
(1996), respectively.
Wiener filter has been applied to the reconstruction of CMB maps
from different experiments, such as COBE (Bunn \etal 1994,1996)
and Saskatoon (Tegmark \etal 1997).
The bottom panels of figure~\ref{wiener} reproduces an example of 
the performance of 
Wiener filter for a CMB simulation of a standard flat CDM model 
plus Gaussian noise with a signal to noise ratio $S/N=1$. 
The level of noise is greatly reduced 
(by a factor $\sim 5$) in the reconstructed image, although the
small scale structure of the signal is also suppressed.

\begin{figure}[!t]
\caption[Reconstructed maps for wavelets and Wiener filter]
{Simulated map of the cosmological signal for a standard CDM 
model (top left), signal plus
Gaussian noise with $S/N = 1$ (top right), denoised map using wavelets 
(middle left), residual map obtained from the CMB signal map 
minus the denoised one (middle right), denoised map using Wiener 
filter (bottom left) and the corresponding residuals (bottom right).}
\label{wiener}
\end{figure}

Methods based on Wiener filter are one of the alternatives to 
reconstruct the cosmological signal as well as the foregrounds from 
the future multifrequency all-sky CMB data (Tegmark \& Efstathiou 1996).

\subsection{Maximum Entropy Method}
\label{mem}
The maximum entropy method (MEM) is derived in the context of the Bayesian
formalism. Bayes' theorem states that
the posterior probability $P(\mathbf{s}|\mathbf{d})$
of an underlying signal $\mathbf{s}$ given some data $\mathbf{d}$
is proportional to:
\begin{equation}
P(\mathbf{s}|\mathbf{d}) \propto P(\mathbf{d}|\mathbf{s})
P(\mathbf{s}) \, ,
\label{mibayes}
\end{equation}
where $P(\mathbf{d}|\mathbf{s})$ is the likelihood (probability of obtaining
a set of data given an underlying signal) and $P(\mathbf{s})$
is the prior probability. The Bayesian estimator $\mathbf{\hat{s}}$ of the 
signal is chosen to be the one that maximises the posterior probability 
given in equation(\ref{mibayes}).
The form of the likelihood is determined from the data. 
For the case of Gaussian noise (assumed to have zero mean for simplicity), 
the likelihood is a N-multivariate Gaussian distribution:
\begin{equation}
P(\mathbf{d}|\mathbf{s}) \propto \exp\left[-\frac{1}{2}
(\mathbf{d}-\mathbf{Rs})^t \mathbf{N}^{-1}(\mathbf{d}-\mathbf{Rs}) \right] 
\propto \exp \left(-\frac{\chi^2}{2}\right) \, . 
\end{equation}

On the other hand, we must assume a prior probability, which includes 
our knowledge from the underlying signal. 
A possible choice is given by MEM. This method 
introduces a conservative prior,
that chooses, from all the possible signals compatible with the
data, the one with less structure. The entropic prior is given
by (Skilling 1989):
\begin{equation}
P(\mathbf{s}) \propto \exp\left[\alpha S(\mathbf{s},\mathbf{m}) \right] \, ,
\end{equation}
where $\mathbf{m}$ is a model vector to which $\mathbf{s}$ defaults 
in the absence 
of data and $\alpha$ is a constant that depends on the scaling of the 
problem. The function $S(\mathbf{s},\mathbf{m})$ is the cross 
entropy of $\mathbf{s}$ and $\mathbf{m}$.
For the case of a positive additive distribution, as is the case 
in standard applications of maximum entropy,  
$S(\mathbf{s},\mathbf{m})$ is given by (Skilling 1989):
\begin{equation}
\label{entropia}
S(\mathbf{s},\mathbf{m})=\sum_{i=1}^M\left(s_i-m_i-s_i ln\left[
\frac{s_i}{m_i}\right] \right) \, .
\end{equation}
However, in the case of the CMB, there are both positive and negative
fluctuations. Therefore, this expression needs to be generalized
for images that take positive and 
negative values. This is done 
by considering the image to be the difference of two positive
additive distributions $\mathbf{s}=\mathbf{u}-\mathbf{v}$ 
(Hobson \& Lasenby 1998), obtaining for $S(\mathbf{s},\mathbf{m})$:
\begin{equation}
\label{sneg}
S(\mathbf{s},\mathbf{m})=\sum_{i=1}^M\left(\psi_i-m_{ui}-m_{vi}
-s_i \ln\left[ \frac{\psi_i+s_i}{2 m_{ui}}  \right] \right) \, ,
\end{equation}
where $\psi_i=\left(s_i^2+4m_{ui}m_{vi} \right)^{1/2}$ and 
$\mathbf{m}_u$ and $\mathbf{m}_v$ are separated models for 
$\mathbf{u}$ and $\mathbf{v}$, respectively.

Thus, the posterior probability can be written as:
\begin{equation}
P(\mathbf{s}|\mathbf{d}) \propto \exp\left(-\frac{\chi^2}{2}+
\alpha S(\mathbf{s},\mathbf{m}) \right) \, .
\end{equation}
Therefore, maximising $P(\mathbf{s}|\mathbf{d})$ is equivalent
to minimise the quantity $\phi=\frac{\chi^2}{2}-\alpha S$.
The constant $\alpha$ can be interpreted as a regularising parameter 
of the relative weight of the data and the prior. One possible
way to determine $\alpha$ is introducing it as an extra parameter
in the Bayesian framework (see Skilling 1989, Hobson \etal 1998a).

Different choices of the prior are possible, resulting in different
reconstructions. In particular, for the case of a Gaussian prior, we 
recover the Wiener filter. 
Actually, MEM and Wiener filter are closely related. 
It can be shown, that in the small fluctuation limit, Wiener filter
can be recovered as a quadratic approximation to MEM
(e.g. Hobson \etal 1998a). For a comparison of the performance of 
both methods see Hobson \etal 1998a.

MEM has been applied to CMB data obtained
from different experiments (White \& Bunn 1995, Jones 1998, 
Jones \etal 1998) as well as to simulations of
the future satellite missions (Hobson \etal 1998a,1999, 
Jones \etal 1999).
In particular, a Fourier-space MEM has been used by the previous authors 
to perform a separation
of the different components of the microwave sky for Planck
simulations of size $10\deg \times 10\deg$. In addition to the
cosmological signal and the instrumental noise, the simulations include
the contribution of the Galactic components (dust, free-free and
synchrotron), extragalactic point sources as well as the thermal and 
kinetic SZ effects from cluster of galaxies. The method performs very
well, producing accurate maps and power spectra of the CMB. 
Moreover, given some prior knowledge of the power spectra of the
CMB and foregrounds, it is also possible to recover accurate
reconstructions of the thermal SZ effect and the Galactic components.
In addition, precise catalogues of point sources can be recovered
at each of the Planck frequency channels.

\subsection{Wavelet techniques}

The development of wavelet techniques applied to signal processing has
been very fast in the last ten years. They are known to be very efficient
in dealing with problems of data compression and denoising and could
be a good alternative to analyse CMB data.
The property that makes wavelets so interesting is that they keep
a good space-frequency localization.
Each point of the signal is associated to a set of wavelet coefficients
corresponding to different scales. Thus, unlike the Fourier transform, 
the wavelet transform allows to have information about the importance of 
different scales at each position.
There is not a unique choice for the wavelet basis.
We will only consider here the 1-dimensional case for 
discrete, orthogonal and compactly-supported wavelet bases.
The basis is constructed from dilations and translations of the 
{\it mother} (or {\it analysing}) wavelet function $\psi$ and a 
second related function $\phi(x)$ called the {\it father} (or 
{\it scaling}) function:
\begin{eqnarray}
\psi_{j,l} & = & 2^{\frac{j-n}{2}}\psi\left( 2^{j-n}x-l\right) \, , 
\nonumber \\
\phi_{j,l} & = & 2^{\frac{j-n}{2}}\phi\left( 2^{j-n}x-l\right) \, , 
\end{eqnarray}
where $ 0 \ge j \ge n-1$ and $ 0 \ge l \ge 2^j-1 $ are integer denoting 
the dilation and translation
indices, respectively, and $2^n$ is the number of pixels of 
the considered discrete signal $f(x)$.
$\psi$ and $\phi$ must together satisfy some mathematical relations, as 
first shown by Daubechies (1988).
In particular, the most straightforward requirements are:
\begin{eqnarray}
\int \psi(x)dx & = & 0 \, , \nonumber \\
\int \phi(x)dx & = & 1 \, . 
\end{eqnarray}
The reconstruction of the signal $f(x)$ using the wavelet basis is 
given by:
\begin{equation}
f(x)= a_{0,0}\phi_{0,0}(x) + \sum_j \sum_l w_{j,l}\psi_{j,l}(x)
\, ,
\end{equation}
being $a$, $w$ the wavelet coefficients defined as:
\begin{eqnarray}
a_{0,0} & = & \int f(x) \phi_{0,0}(x) dx \, , \nonumber \\
w_{j,l} & = & \int f(x) \psi_{j,l}(x) dx \, .
\end{eqnarray}

Denoising of data using wavelets are based on the different 
scale properties of noise and signal.
The idea is to keep those coefficients dominated by the signal and set
to zero those where the noise is the main contribution.
This can be achieved by 
using thresholding techniques. Given a set of data
$\mathbf{d}$, we can recover the underlying signal by acting over the data
wavelet coefficients $w_{j,l}$ in the following way:
\begin{equation}
\hat{w}_{j,l} = \left\{ \begin{array}{ll}
w_{j,l}-\nu \sigma_n & {\rm if~} w_{j,l} > \nu \sigma_n \\ 
0 & {\rm if~}  |w_{j,l}| \le \nu \sigma_n \\
w_{j,l}+\nu \sigma_n & {\rm if~} w_{j,l} < -\nu \sigma_n \\ 
\end{array} \right.
\end{equation}

This is known as a {\it soft} threshold (Donoho \& Jonhstone 1995).
$\nu$ is the threshold defined
in units of the dispersion of the noise $\sigma_n$ at each scale. 
For the case of orthogonal wavelets and uncorrelated noise,
$\sigma_n$ is constant over all the different scales. 
By inverse transforming the thresholded coefficients we get an estimation of
the underlying signal, where the noise has been highly suppressed.
The choice of the threshold can be made using signal-independent
prescriptions such as the Stein's Unbiased Risk Estimate (SURE)
(e.g. Ogden 1997).

Wavelet techniques have been applied to very different fields
in the last years. However, only very recently some works have studied 
their performance when dealing with CMB maps (Sanz \etal 1999a,b,
Tenorio \etal 1999). The middle panels of Figure~\ref{wiener}
show the reconstructed image obtained by applying a wavelet 
thresholding technique, as discussed by Sanz \etal 1999b.
Even although knowledge of the power spectrum of the original
signal is not required for this technique, the noise is suppressed as
efficiently as in the Wiener filter case. In addition, the
power spectrum is recovered up to $\ell \simlt 1700$ for $S/N \le 1$ with
an error $\simlt 20 \%$ what is considerably better than the power
spectrum of the reconstructed image provided by Wiener filter.
However, the non-linearity of the soft thresholding technique is
introducing a certain level of non-Gaussianity, what must be taken
into account when analysing the data. 
This application considers only the presence of the cosmological signal
plus Gaussian noise, being just a first approach that pretends 
to shed light on the wavelet 
characterization of the different components. The final goal would
be a Bayesian framework (incorporating entropy or other constraints)
dealing with wavelet components at different scales and introducing
multifrequency information. Following this direction, some work has
already been carried out by Jewell \etal 1999.

In addition to denoising, wavelet techniques are good in detecting 
structure. For instance, the mexican hat wavelet (a non-orthogonal
wavelet) seems to be a good tool to detect and subtract point sources
from the CMB (Cay\'on \etal 1999).

\section{Conclusions}

In this paper, we have reviewed the present status of CMB experiments, 
summarize
some basics aspects of the theory, describe some relevant characteristics
of the foregrounds, review the methods proposed to test Gaussianity and 
outline the reconstruction methods recently applied to CMB.

During the last years there has been an explosion of experiments
dedicated to measuring the CMB anisotropies, including balloon-borne
and ground based instruments. More than a dozen groups have reported
the detection of CMB anisotropies. In addition two satellite missions,
MAP and Planck, will provide with multifrequency all-sky CMB data with
unprecedented resolution and sensitivity.
In order to obtain all the valuable information encoded in the CMB, a good 
understanding of the underlying theory is necessary and we have 
reviewed the basics of primary and secondary anisotropies.
On the other hand, our ability to measure the cosmological parameters will
also depend on the removal of the Galactic and extragalactic foregrounds.
We have reviewed the current knowledge of these
contaminants, pointing out the presence of an anomalous component.
Further study of these foregrounds becomes necessary in order to
analyse successfully the CMB data.
One of the key issues expected to be solved 
with the future CMB data is 
whether the temperature anisotropy field is Gaussian as predicted
by the standard inflationary model.
Numerous tests have been proposed to detect non-Gaussianity in the CMB, 
some of them looking very promising.
However, before any general conclusion can be established, a direct
comparison of the different methods is necessary.
These tests, computed from the data in a manner completely independent
of the $\cl's$, also provide a useful check on the 
conclusions derived from the power spectrum, even in the case of 
underlying Gaussian fluctuations.
Finally, we have briefly described the basics of reconstruction 
methods recently applied to the CMB, including Wiener filter, 
the maximum entropy method and wavelet techniques.
The development of an integrated scheme combining the power of Bayesian 
methods and the excellent properties of wavelet techniques
is one of the promising
projects left for the future in order to separate and reconstruct
the different components of the microwave sky.

\ack
RBB thanks Jos\'e Luis Sanz and Enrique Mart\'\i nez-Gonz\'alez
for a careful reading of the manuscript,
Laura Cay\'on and Patricio Vielva for useful comments and
Max Tegmark for kindly providing one of the figures.
This work has been supported by the DGESIC Project no. PB95-1132-C02-02, 
CICYT Acci\'on Especial no. ESP98-1545-E and Comisi\'on Conjunta 
Hispano-Norteamericana de Cooperaci\'on Cient\'\i fica y Tecnol\'ogica
with ref. 98138. The author also acknowledges financial support from a 
Spanish MEC fellowship and from the PPARC in the form of a research
grant.

\bigskip
{\bf CMB experiments}
\begin{description}
\item[ACBAR:] http://cfpa.berkeley.edu/$\sim$swlh/research/acbar.html
\item[ACE:] http://www.deepspace.ucsb.edu/research/Sphome.htm
\item[APACHE:] http://tonno.tesre.bo.cnr.it/$\sim$valenzia/APACHE/apache.htm
\item[Archeops:] http://www-crtbt.polycnrs-gre.fr/archeops/Egeneral.html
\item[ARGO:] http://oberon.roma1.infn.it/argo.htm
\item[ATCA:] http://www.atnf.csiro.au/research/cmbr/cmbr\verb+_+atca.html
\item[BAM:] http://cmbr.physics.ubc.ca/experimental.html
\item[BEAST:] http://www.deepspace.ucsb.edu/research/Sphome.htm
\item[BOOMERanG:] http://astro.caltech.edu/$\sim$lgg/boom/boom.html
\item[CAT:] http://www.mrao.cam.ac.uk/telescopes/cat/index.html
\item[CBI:] http://phobos.caltech.edu/$\sim$tjp/CBI/
\item[CG:] http://brown.nord.nw.ru/CG/CG.htm
\item[COBE:] http://www.gsfc.nasa.gov/astro/cobe/cobe\verb+_+home.html
\item[DASI:] http://astro.uchicago.edu/dasi/
\item[FIRS:] http://pupgg.princeton.edu/$\sim$cmb/firs.html
\item[HACME/SP:] http://www.deepspace.ucsb.edu/research/Sphome.htm
\item[MAP:]http://map.gsfc.nasa.gov
\item[MAT:] http://imogen.princeton.edu/$\sim$page/matdir/www/index.html
\item[MAX:] http://cfpa.berkeley.edu/group/cmb/gen.html
\item[MAXIMA:] http://cfpa.berkeley.edu/group/cmb/gen.html
\item[MSAM:] http://topweb.gsfc.nasa.gov/
\item[OVRO:] http://www.cco.caltech.edu/$\sim$emleitch/ovro/ovro\verb+_+cmb.html
\item[Planck:] http://astro.estec.esa.nl/Planck
\item[POLAR:] http://cmb.physics.wisc.edu/polar/
\item[Polatron:] http://phobos.caltech.edu/$\sim$lgg/polatron/polatron.html
\item[PYTHON:] http://cmbr.phys.cmu.edu/pyth.html
\item[QMAP:] http://pupgg.princeton.edu/$\sim$cmb/qmap/qmap.html
\item[Saskatoon:]http://pupgg.princeton.edu/$\sim$cmb/skintro/sask\verb+_+intro.html
\item[SPort:] http://tonno.tesre.bo.cnr.it/$\sim$stefano/sp\verb+_+draft.html
\item[SuZIE:] http://phobos.caltech.edu/$\sim$lgg/suzie/suzie.html
\item[Tenerife:] http://clarin.ll.iac.es/
\item[TopHat:] http://topweb.gsfc.nasa.gov/
\item[Viper:] http://cmbr.phys.cmu.edu/vip.html
\item[VLA:] http://www.nrao.edu/vla/html/VLAhome.shtml
\item[VSA:] http://www.mrao.cam.ac.uk/telescopes/vsa/index.html
\end{description}

\end{document}